\documentclass[useAMS,usenatbib,referee]{mn2e}

\usepackage{graphicx      } 
\renewcommand{\vec}[1]{\mbox{\boldmath $#1$}} 

\title[Particle Simulation for the Global Pulsar Magnetosphere]%
{A Particle Simulation for the Global Pulsar Magnetosphere:
the Pulsar Wind linked to the Outer Gaps}
\author[T. Wada and S. Shibata]{%
Tomohide Wada$^1$\thanks{E-mail: wada@ksirius.kj.yamagata-u.ac.jp}
\& Shinpei Shibata$^2$\thanks{E-mail: shibata@sci.kj.yamagata-u.ac.jp}%
\thanks{This file has been amended to
highlight the proper use of \LaTeXe\ code with the class file.
These changes are for illustrative purposes and do not reflect the
original paper by A. V. Raveendran.}\\
$^1$School of Science and Technology, Yamagata University,
Kojirakawa, Yamagata 990-8560, Japan \\
$^2$Department of Physics, Yamagata University,
Kojirakawa, Yamagata 990-8560, Japan}
\begin{document}
\date{Accepted 2006 December 22. Received 2006 December 9; 
in original form 2006 June 24}

\pagerange{\pageref{firstpage}--\pageref{lastpage}} \pubyear{2006}

\maketitle

\label{firstpage}

\begin{abstract}
Soon after the discovery of radio pulsars in 1967,
the pulsars are identified as strongly magnetic (typically
$10^{12}$G) rapidly rotating ($\sim 10^{2}-0.1$ Hz) neutron
stars. However, the mechanism of particle acceleration in the pulsar
magnetosphere has been a longstanding problem.
The central problem is why the rotation power manifests itself in both
gamma-ray beams and
a highly relativistic wind of electron-positron plasmas, which excites
surrounding nebulae observed in X-ray.
Here we show with a three dimensional particle simulation
for the global axisymmetric magnetosphere that
a steady outflow of electron-positron pairs is formed
with associated pair sources,
which are the gamma-ray emitting regions within
the light cylinder.
The magnetic field is assumed to be dipole, and to be consistent,
pair creation rate is taken to be small, so that the model might be
applicable to old pulsars such as Geminga.
The pair sources are charge-deficient regions around the null
surface, and we identify them as the outer gap.
The wind mechanism is the electromagnetic induction
which brings about fast azimuthal motion
and eventually trans-field drift by radiation drag
in close vicinity of the light cylinder and beyond.
The wind causes loss of particles from the system.
This maintains charge deficiency in the outer gap and pair creation.
The model is thus in a steady state, balancing loss and supply of particles.
Our simulation implies how the wind coexists with
the gamma-ray emitting regions in the pulsar magnetosphere.
\end{abstract}


\section{Introduction}

Our particle simulation follows 
the historic gedanken experiments 
by \citet{b1} 
and by \citet{b2}, 
in which a rotating neutron star with aligned magnetic
moment is put in vacuum, 
and motion of charged particles around the star are explored.
Electromagnetic induction by the star produces 
a quadrupolar electric field in the surrounding
space with a surface charge on the star.
The electric force is so large that
the surface charge is pulled out
into the magnetosphere.
For the fate of these charged particles,
it was shown by the earlier simulations
(Krause-Polstorff \& Michel 1985a; 1985b, hereafter KM)
that the magnetosphere settles down into a quiet state with
electronic domes above the polar caps, 
a positronic or ionic equatorial disc and
vacuum gaps in the middle latitudes.
(For definite of sign of charge, 
we consider a parallel rotator
rather than an anti-parallel rotator; with this polarity 
the polar caps are negatively
charged and the equatorial region is positively charged.)
The quiet solution has no energy release at all.
However, the gaps are unstable against pair creation cascade,
which is the subject of our simulation for an active magnetosphere.

Before describing the results of our simulation, let us consider
how plasma density and dissipation processes affect the magnetic
filed structure.
Here, we restrict ourselves to the axisymmetric case for which
many works have been done.
In the limit of low density, the electromagnetic fields around
the star is vacuum ones, and the magnetic field lines are closed.
As has been shown by the previous particle simulation (KM),
for a moderate density of plasmas,
the electromotive force by the star manifests itself in strong charge
separation. The magnetosphere contains the gaps in between 
the polar domes and the equatorial disc.
It is known that convective current by the corotating clouds 
with finite extent
is not enough to open the magnetic field
(KM, Kaburaki 1982, Fitzpatrick \& Mestel 1988).

To obtain the quiet solutions with finite extent of plasmas
in space,
the total charge of the system $Q$ is assumed to be non-zero.
If $Q=0$, the corotating plasma will extend to the 
light cylinder and possibly beyond the light cylinder (KM). 
However, such a solution has yet to be discussed
seriously.
Rylov (1978) and Fitzpatrick \& Mestel (1988) considered
a flow beyond the light cylinder in closed magnetic filed lines.
In this case, non-electromagnetic force such as radiation drag and
inertial force is required for a self-consistent steady solution.
The reaction force due to rotational bremsstrahlung may be
important: 
the radiation reaction force becomes comparable to the Lorentz
force when
$\gamma \sim (3 B R_c^2/2 q)^{1/4} \equiv \gamma_r$,
where $B$ is the magnetic field strength,
$R_c$ is the curvature radius, and $q$ is the charge of a particle.
A typical value of $\gamma_r$ is $\sim 3 \times 10^7 (\mu_{30}/P)^{1/4}$
for electrons,
where $P$ is the rotation period in seconds and $\mu_{30}$ is the magnetic
moment in $10^{30} {\rm G} \; {\rm cm}^3$.
The value of $\gamma_r$ is still smaller in a region
where the magnetic field is weaken in vicinity of the 
equatorial neutral sheets.
The expected global flow pattern is circular, being similar to the electric 
quadrupolar fields, starting from the star, 
going beyond the light cylinder and returning back to the star. 
A possibility is also pointed out that both positive and negative
particles may leave the star to make a wind toward infinity 
in the steady state (Jackson 1976, Fitzpatrick, \& Mestel 1988).

Contrary to the case of moderate densities, the force-free model
assumes high density neutral plasmas and imposes the boundary conditions 
that the magnetic field is opened. However,
the appropriate boundary condition for the force-free problem is still
controversial (Uzdensky 2003).
It has been shown that 
the separatrix, which is the open-close boundary, should not have
a sheet current if the Y-point  
is located on the light cylinder, and that
a wedge-shaped electric-field-dominant region ($E>B$) appears beyond
the light cylinder (Uzdensky 2003).
Ogura and Kojima (2003) obtains a numerical solution of the force-free 
magnetosphere, and find that electric field becomes larger than the
magnetic field in a region beyond a few light-cylinder radii,
suggesting break down of the force-free approximation.
Relativistic magnetohydrodynamic simulation also requires some dissipation
in vicinity of the magnetic neutral sheet, where magnetic field lines are
closed (Komissarov 2006).
Thus, the force-free model suggests that most part of the magnetic
flux will be open, but some flux might be closed with dissipation in
the equatorial neutral sheet.

In any case, the force-free approximation requires sources of copious
quasi-neutral plasmas.
Young pulsars such as Crab will have a high rate of pair creation, and
therefore the open filed structure may be plausible in a large part
of the magnetosphere. On the other hand,
for older pulsars such as Geminga, the pair creation rate is lower, and
therefore a larger gap and closed field structure may be plausible.
Our study starts with a simulation for the latter low-density case
in which the magnetic filed is assumed to be dipole.
To be consistent, pair creation rate is assumed to be low.
Cases with higher densities shall be treated in subsequent papers.
Our result will be helpful also for understanding the electrodynamics in the
dissipative domain with closed magnetic fields.


\section{Method of Simulation}

In the first step of our calculation, 
we have reproduced the quiet solution. However, it must be noted here that
we use a distinctively different method.
In the previous works, to reduce load in calculation,
the particles are assumed to follow magnetic field lines and 
to obey the Aristotelian mechanics, i.e., the particles are put at rest if the
exerting force vanishes. 
We solve the relativistic equation of motion
for each particle like particle-in-cell (PIC) methods, so
that we treat gyro-motion and any kinds of drift motions.
The mass-charge ratio of the simulation super-particles is taken to be so small that gyro-motion
is kept always microscopic.
In the case of actual pulsars,
gyro-motion can be macroscopic if the Lorentz factor $\gamma$ is increased up to
the maximum reachable value $\gamma_{\rm max}$,
which corresponds to the available voltage of the star, in other words, the
potential drop across the polar caps.
However, this would not be the case because not all the available voltage
is used for field-aligned acceleration in the gaps, and radiation drag 
prevents $\gamma$ from increasing. 
In our simulation, typical gyro-radii 
for the light cylinder magnetic field
are $\sim 5\times 10^{-4} \gamma R$,
where $\gamma$ is the Lorentz factor of the super-particles, and
$R$ is the stellar radius.
The value of $\gamma_{\rm max}$ for the super-particles is 2000.
In the simulation, $\gamma \ll \gamma_{\rm max}$, so that gyro-radii
are always much smaller than $R$.
As mentioned before, 
we take the radiation drag force into account
in our equation of motion, which becomes
\begin{equation}
{d \vec{p}_i \over dt} = q_i ( \vec{E} + \vec{\beta}_i \times 
\vec{B} ) + \vec{F}_{r},
\end{equation}
where $\vec{\beta}_i$ is the velocity of the $i$-th particle in
units of the light speed; $\vec{p}_i = m \gamma_i \vec{\beta}_i$,
$\gamma_i = (1- \vec{\beta}_i \cdot \vec{\beta}_i )^{-1/2}$;
$m$ and $q_i$ ($=+q$ or $-q$) is the mass and charge of the particles;
$\vec{F}_r = (2/3)(q_i^2/R_c^2) \gamma_i^4 (\vec{p}_i / |\vec{p}_i|)$
represents the radiation drag force.

Unlike typical PIC,
the electric field is calculated in use of a Green's function
with the boundary condition that the stellar surface
is a perfect conductor. 
The method enables us to keep exactly the
corotational potential on the stellar surface.
We calculate the electric field and particle motion in three dimension,
and therefore the scheme itself is fully three dimensional.
The model we considered is axisymmetric
because of the boundary condition of aligned rotator. 
The electric field is composed of the vacuum part $\vec{E}_v$,
which is produced by a rotating star in vacuum,
and the particle part $\vec{E}_m$, which is produced by particles
in the magnetosphere:
\begin{equation}
\vec{E}(\vec{r}) = \vec{E}_v(\vec{r}) + \vec{E}_m(\vec{r}) 
= - \nabla \phi_v - \nabla \phi_m
\end{equation}
with
\begin{equation}
\phi_v = \left\{ \begin{array}{lr}
\displaystyle
{B_s R \; \Omega \over 6} \left[ 3 \left(r \over R \right)^2
\sin^2 \theta - 2 \right], 
& (r \lid  R) \\
\displaystyle
B_s R \left[ - {\Omega \over 6} \left( R \over r \right)^3 
(3 \cos^2 \theta - 1 ) + Q {R \over r} \right],
& (r > R)
\end{array} \right.
\end{equation}
and
\begin{equation}
\phi_m = \sum^{n}_{i=1} q_i \left[
{1 \over |\vec{r} - \vec{r}_i |}
- {R/r_i \over |\vec{r} - (R/r_i)^2 \vec{r}_i |}
- {1 - R/r_i \over |\vec{r}|} \right],
\label{GF}
\end{equation}
which yields
\begin{equation}
\vec{E}_m (\vec{r})
  = \sum_{i=1}^{n} q_i \left[ 
 {\vec{r}-\vec{r}_i \over |\vec{r}-\vec{r}_i|^3}
- \frac{R}{r_i}
{\vec{r}-(R/r_i)^2\vec{r}_i \over |\vec{r}-(R/r_i)^2\vec{r}_i|^3}
- \frac{(1-R/r_i) \vec{r} }{r^3}
 \right],
\end{equation}
where $\vec{r}$ is the position vector for a given point,
$q_i$ and $\vec{r}_i$ are the charge and the position of the $i$-th particle,
and $n$ is the particle number;
$B_s$ is the magnetic field strength at the poles,
$\Omega = R/R_L $ is the non-dimensional angular velocity, and
$Q$ is the non-dimensional net-charge in units of $B_s R^2$.
Freedom of the net-charge of the system $Q$ is included in $\vec{E}_v$.
The Green's function,  the brackets in (\ref{GF}), 
is composed of Coulomb potentials by the particles in the magnetosphere and
their mirror charges, so that $\phi_m$ satisfies the boundary conditions,
$\phi_m =$ constant on the stellar surface and $\phi_m \rightarrow 0$ at infinity.
Since $\vec{E}_v$ represents the corotation electric field by the star,
$\vec{E}_v + \vec{E}_m$, 
is automatically guaranteed to be the corotational electric field on the stellar surface.
In general, surface charge appears, and field-aligned
electric force exerts on it. 
We, therefore,  replace the surface charge by simulation particles
in each step of time, 
so that the surface charge can move into the 
magnetosphere freely.

In (\ref{GF}), we neglect the retardation effect.
By this neglect, the calculation of the electric field
becomes simply summing-up of Coulomb forces, which
can be done very quickly by the special purpose computer for
astronomical $N$-body problem,
GRAPE (Sugimoto et al. 1990; Makino et al. 1997). 
Typical use of GRAPE is for stellar dynamics.
Note that GRAPE has sign-bits and 
applies to the Coulomb forces.
Omission of retardation is not serious as far as
an axisymmetric and steady state is concerned, while
introduction of GRAPE makes us possible to perform this heavy
calculation.

For realistic simulation of the active pulsars, 
we take into account electron-positron pair creation.
If there is a very large electric field along the magnetic field,
particles are accelerated to emit curvature gamma-rays which converted
into pairs by interaction with soft photons from the stellar surface
or with strong magnetic field near the star.
In our simulation, we omit detailed pair creation processes and
assume that
pairs are created if a region has a  field-aligned electric field larger than
a critical value $E_{cr}$. For the results presented in this paper,
$E_{cr} = 0.25 B_L$, where $B_L$ is the light cylinder magnetic field.
The creation rate is a simulation parameter and limited by ability of 
computer. In the present calculation, typically $1600$ particles are 
created per rotation. 
The total number of particles is
$\sim  5 \times 10^4$ in a steady state.

As prescribed by our choice of the Green's function, 
the boundary condition for the electric potential has already
been included, i.e., the corotational potential 
(or $E_\theta = B_s \Omega \sin \theta \cos \theta$) on the 
stellar surface and $\phi \rightarrow 0$ at infinity.
It is a distances of about $10 R_L$ that
the electric potential vanishes within the accuracy of the calculation,
which is $1.5 \times 10^{-2} B_L R_L$ in our simulation.
While the outer boundary condition for the electric filed is given 
at infinity,
we put an outer boundary for particles at $R_{out}=10R_L$, so that
the particles flowing out across the outer boundary are removed, 
and no particles come into the simulation domain across it,
and thereby we save the number of particles.
We also used $R_{out} = 20R_L$, and $30R_L$ to check the effect of
the outer boundary. 
For the initial condition, we started with a vacuum magnetosphere, 
and suppress pair creation 
until the static solution by KM is reached. Once we establish the static
solution, then pair creation is switched on.
The simulation parameters are $\Omega =R/R_L =0.2$,
$m/q = 10^{-5}$, and 
$B_L =(1/2)\Omega^3 B_s =4 \times 10^{-3} B_s$.

\section{Results and Discussion}

\begin{figure}
\begin{center}
%
%
\parbox{170mm}{
\includegraphics[height=80mm]{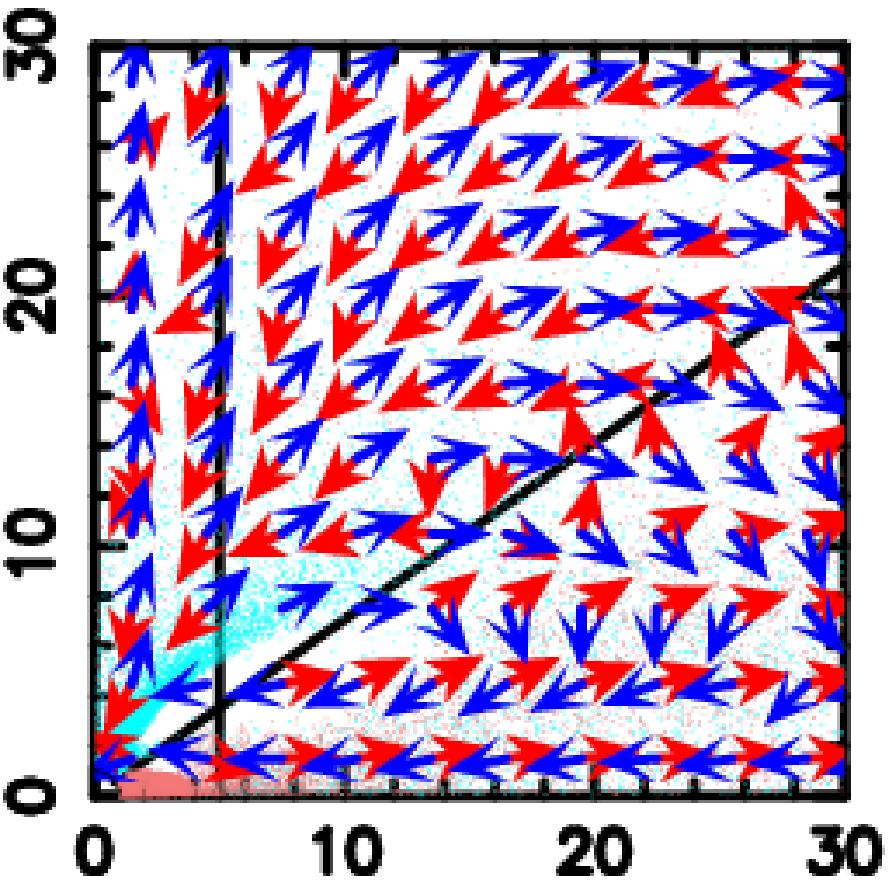} \hspace{5mm}
\includegraphics[height=80mm]{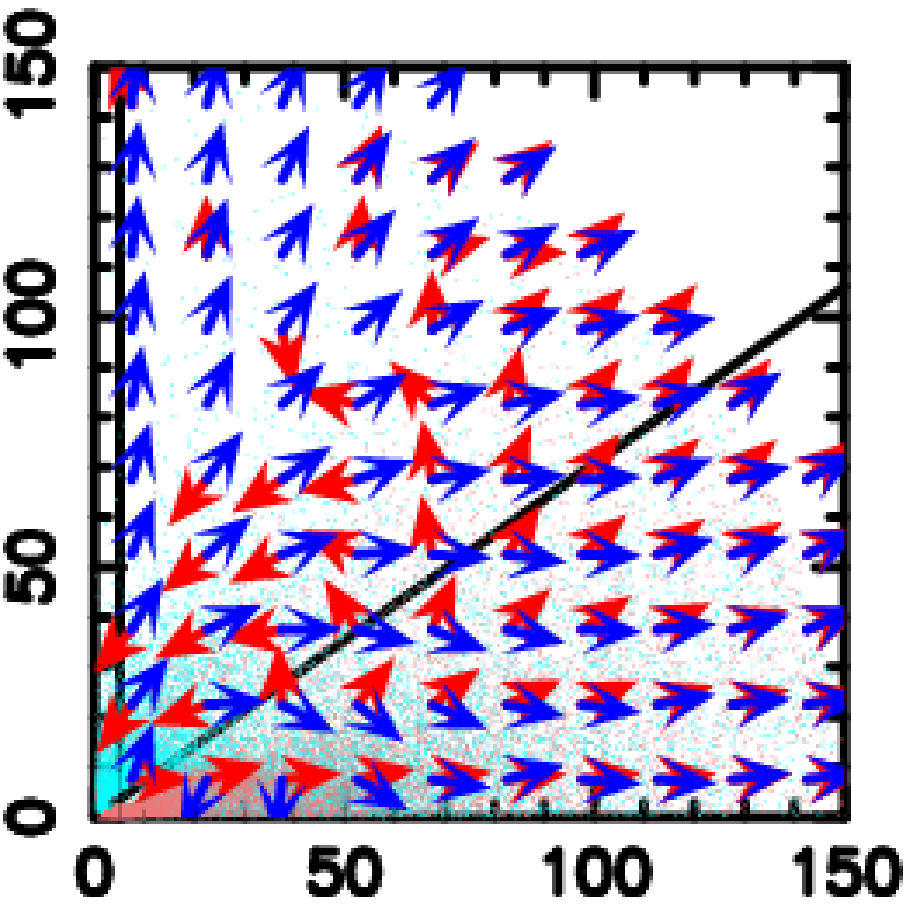} \\
\includegraphics[height=80mm]{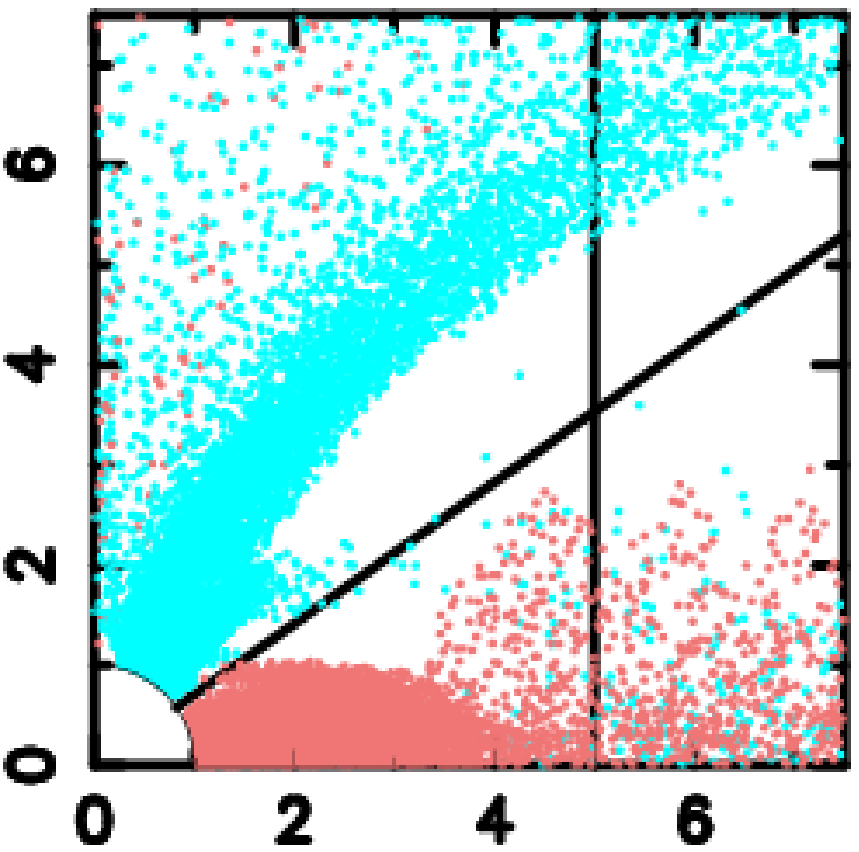} \hspace{5mm}
\includegraphics[height=80mm]{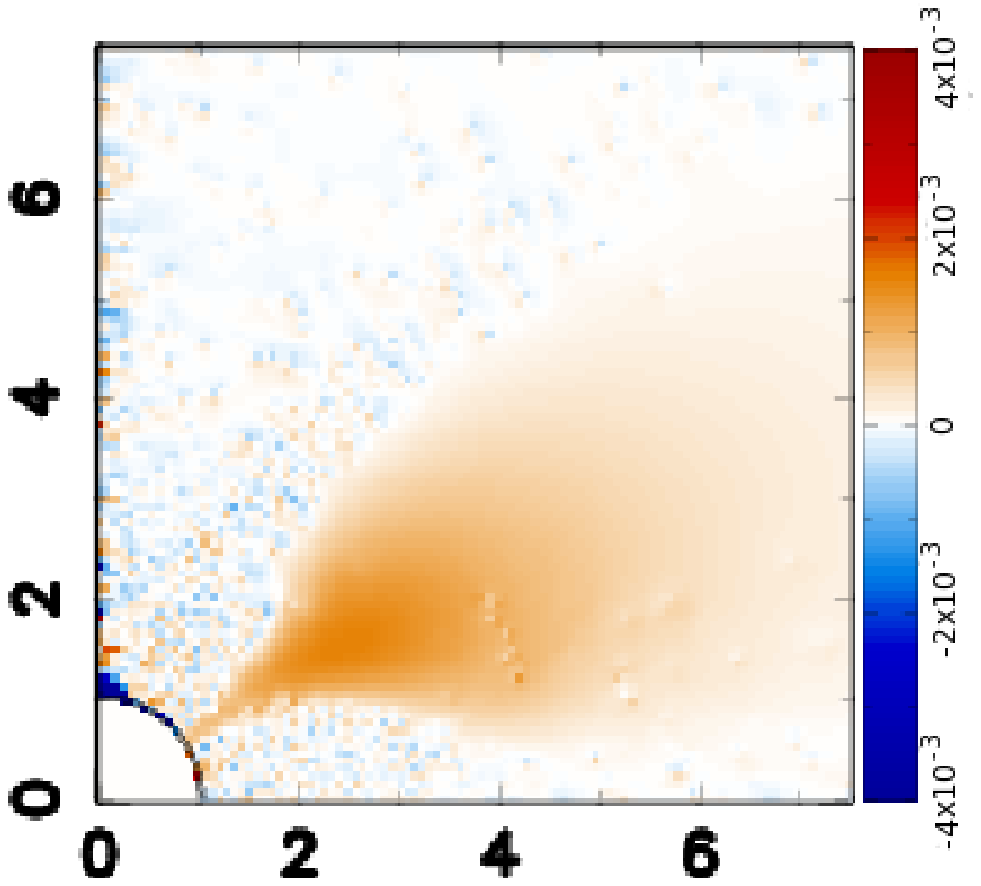} \\
}
\caption{ \label{fig1} %
{\it Top left panel}: 
The velocity fields (arrows) and the positions of the particles
(dots) on the meridional plane.
The red arrows and light red  dots are for positive particles, 
and the blue arrows and light blue dots are for negative particles.
The geometrical scale length is normalized by the stellar radius, and
the light-cylinder corresponds to the axial distance of 5.
The inclined line in the middle latitude indicates the null 
surface for the dipole field, on which the Goldreich-Julian charge 
density vanishes.
{\it Top right panel}: 
The same, but for the outer magnetosphere.
{\it Bottom left panel}: 
The outer gap appears
at axial distance of $\sim 3$ around the null surface,
where the positive particles are ejected to the right, 
and the negative particles to the left.
Only the particle positions are shown.
{\it Bottom right panel}: 
The strength of the field-aligned electric fields.
}
\end{center}
\end{figure}

After several rotation-periods of time, the magnetosphere
settles into a steady state with electronic polar outflows and
positronic equatorial outflows, which are mixed in
the middle latitudes (see the top panels of Fig.~\ref{fig1}).
Some particles are circulating in the magnetosphere.
However, the rest of the particles forms the wind.
A region around the null surface produces electron-positron pairs
steadily, and the accelerating electric field is kept slightly larger 
than $E_{\rm cr}$, typically $\sim 1.5 E_{\rm cr}$.
In the course to the steady state,
after the onset of pair creation,
the gaps diminish both in size and in electric field strength:
the original gap field is screened out
down to $E_{cr}$, except for the geometrically small
gaps locating around the null surface at axial distances 
of $\sim 0.5 R_L$ ( the bottom panels of Fig.~\ref{fig1}b).
This region is identified as the outer gap
(Cheng, Ho \& Ruderman 1986; Takata, Shibata \& Hirotani 2004) 
which has been proposed to explain the gamma-ray beams.
The bottom right panel of Fig.~\ref{fig1} shows that 
the field-aligned electric field
is unscreened in the gaps. On the other hand, 
the outside of the gaps 
as well as inside the polar domes and the equatorial disc,
the plasmas screen out the field-aligned electric field.

\begin{figure}
\begin{center}
%
\includegraphics[width=73mm]{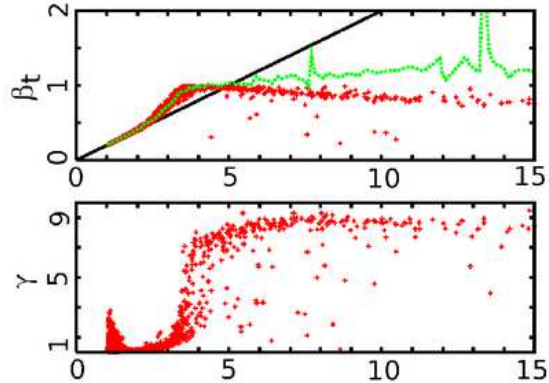}
\caption{ \label{fig2} %
The azimuthal velocity of the particles in the disc ($|z| \leq 0.2$)
(upper panel) and the corresponding Lorentz factor (lower panel) as functions
of the axial distance in units of the stellar radius.
The light radius is located at the axial distance of $\varpi = 5$,
and the Lorentz factor is found to increase around $\varpi \sim 4$.
The dashed green curve indicates the $E_\perp$-drift velocity,
$E_\perp /B$, calculated
formally from the local fields.
The saturation Lorentz factor by the drag is
$\gamma_r \sim 14$ for the super particle
(corresponding to $\gamma_r = 10^4 - 10^5$ for real electrons)
and is 4\% of $\gamma_{\rm max}$ 
}
\end{center}
\end{figure}

The positrons ejected from the outer gap accumulate
in the equatorial disc which grows in size with time.
Due to strong induction of rotation,
the azimuthal velocity of the disc plasma increases
with distance toward the speed of light, and
the Lorentz factor is found to increase until radiation
reaction becomes important.
Since the radiation drag is in azimuthal direction,
the drift across the closed magnetic field causes 
the radial outflow of the positrons (Fig.~\ref{fig1}).
Interestingly the global electric field becomes such
that the equatorial disc {\it super-corotates} as shown in Fig.~\ref{fig2}, 
and as a result, 
radiation drag becomes efficient before the very vicinity of
the light-cylinder, $\sim 0.8 R_L$.
Because the azimuthal motion is due to $E_\perp$ cross $B$ drift,
the perpendicular electric field $E_\perp$ plays an essential role
for the wind mechanism.
Of the positrons drifting across the field,
about one half flows out of the system, 
while the other half drifts to higher latitudes
and turns back to the star.

\begin{figure}
\begin{center}
%
\includegraphics[width=73mm]{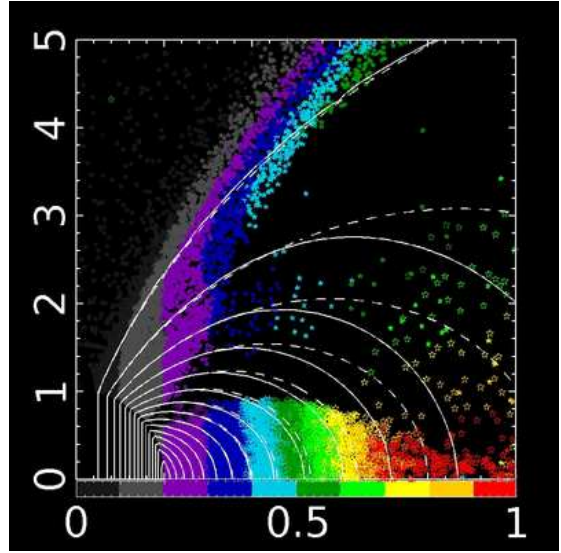} 
\caption{ \label{fig3} %
The azimuthal velocity (color coded) of the particles (dots) and
the contours of the electric potential (dashed-curves)
with the fields lines (solid curves).
The meaning of the colors is shown in the color bar 
just below the equatorial
plane, where each color indicates the corotation speed
at each axial distance.
}
\end{center}
\end{figure}

On the other hand, electrons produced in the outer gap
are absorbed by the star.
However, the equivalent amount is re-emitted 
from the polar caps at higher latitudes.
As a result, the electron domes grow and 
expand across the light cylinder.
In the steady state,
the property of the polar domes changes spatially from the corotating parts
near the star to the sub-rotating parts in higher altitudes as shown
in Fig.~\ref{fig3}, where the azimuthal velocity is indicated by colors
with the electric iso-potentials. The boundaries of the colors are vertical
in the corotating parts, while they bend in the sub-rotating parts, in which
field-aligned motion becomes appreciable.
In the electron clouds extending further out with axial distances of
$\sim 1.5R_L$, increase of the azimuthal velocity ($E_\perp$-acceleration)
causes increase of the Lorentz factor and the drag force.
About one half of the polar electrons flows out of the system 
nearly along the magnetic field lines (the centrifugal outflow), while
the other half drifts across the field lines significantly due to radiation
drag, and
turns back to the star.

Concerning the electric property of the magnetosphere,
we find it most remarkable that the
net charge $Q$ of the system (the star plus magnetosphere) almost vanishes.
In the previous works (KM; Smith, Michel \& Thacker 2001),
a positive $Q$, e.g. our initial value $Q=1/3$ 
(equivalent to $+10$ of KM with their choice of the unit charge), 
is assumed to obtain a finite extent of the plasma clouds, 
and it was suggested that if $Q \sim 0$, the clouds may expand to
the light cylinder.
We have started with a positive $Q=1/3$. 
It is notable that our simulation includes creation and loss of particles,
and the dynamics determines the stable value of $Q$.
In the final steady state,
the positive $Q$ disappears.
More precisely, $Q$ is kept slightly negative so as for electrons
to be pushed and escape.
The net charge $Q$ is adjusted so that the same amounts of
the negative and positive particles flow out steadily.

To see that the particles which are removed at the outer
boundary really leave from the star, we made a convergency study
with $R_{out} = 10R_L$, $20R_L$, and $30R_L$.
When $R_{out}$ is increased,
it becomes much clear that the separatrics between circulating (closed) stream 
lines and wind (open) stream lines is located 
at about $R \sim 8 R_L$ for electrons and 
this position is independent of $R_{out}$.
The loss rate of particle times $R_{out}^2$ increases with $R_{out}$.
Kinetic energies of the removed particles are an order of magnitude
larger than potential energy differences to infinity when $R_{out}
\ga 20R_L$.
These facts indicate
that the particles which reach the outer boundary ($\ga 10R_L$) 
are most likely to reach infinity and to form the wind.
It is also confirmed by the convergency study that the inner 
electric field structure within a few light cylinder radii is not 
affected by $R_{out} \ga 10R_L$.

There is a potential drop by artifact between
the mathematical stellar surface $|\vec{r}| = R$ and 
the bottom of the magnetosphere $|\vec{r}| = R+\Delta$. 
Geometrical thickness of this gap is $\Delta \sim 0.04R$ in our simulation.
The potential gap appears for all
latitudes and is, respectively, negative and positive at high
and low latitudes with magnitude of 
$\sim 0.01 B_L R_L$.
The potential drop is just an artifact due to the fact that
continuous surface charge on the stellar surface is represented
by a finite number of mirror charges.
This effect is similar to that of
work function on the stellar surface.
However, gaps just above the polar caps ($\theta \la 10^\circ$) is
an exception: the potential gap is three times larger than the artifact level.
This ``polar gap'' might be related to dynamics of the polar outflow, and should
obviously be the future interest.

Finally, we calculate change in the magnetic field 
as a perturbation by the obtained electric current. 
Since our simulation is performed for a low density class,
the change obtained is little to be consistent with the assumption
of a dipole field. However, we can detect in the perturbed filed
that the field lines are
pushed outward slightly, and 
that a small fraction of the polar field is opened.
The polar electrons would flow out through the open magnetic flux
in a self-consistent calculation with higher densities.
In a subsequent paper, we would like to treat such cases with
higher pair densities and a finite open magnetic flux by
improving the numerical code.

Observationally, the principal activities of the pulsar magnetosphere
are the pulsar wind and the gamma-ray pulses.
The pulsar powered nebulae observed in X-ray strongly
suggest that the pulsar wind is a relativistic outflow of
pair plasmas.
On the other hand, it is known that
the outer gap well explains the pulsed high energy radiation
(Romani \& Yadigaroglu 1995; Dyks \& Rudak 2003). 
Our simulation shows that
the pulsar wind and the outer gap coexists in a self-consistent
manner.
Recently, the force-free solutions for the axisymmetric magnetosphere
is investigated by several authors (%
Uzdensky 2003; 
Komissarov 2006; 
McKinney 2006%
)
with the assumption of copious pair supply.
One of the controversial topics is dissipation in the magnetic
neutral sheets in the equatorial plane and 
the Y-point at the boundary of open and
close magnetic filed lines.
The azimuthal $E_\perp$ acceleration and radiation drag
may play an important role in the
dynamics of such regions.
In our simulation, these dissipative effects
appear in an exaggerated manner because the particle density is small. 
However, future simulation with higher densities and with
modification of magnetic field should enlighten the dissipation processes
around the light cylinder.

\section*{Acknowledgments}
This work was supported in part by
the National Observatory of Japan for the GRAPE system
(g04b07, g05b04, g06b11) and also by
Grant-in-Aid for Scientific
Research from the Ministry of Education, Culture, Sports, Science and
Technology of Japan (15540227).

\newcommand{\apj}{{ApJ}}
\newcommand{\apss}{{Astrophys. Space Sci.}}
\newcommand{\aap}{{A\&A}}
\newcommand{\mnras}{{MNRAS}}
\newcommand{\nature}{{Nature}}

\label{lastpage}

\end{document}